\documentclass[pre, 12 pt]{revtex4}
\usepackage[cp1251]{inputenc}
\usepackage[english,russian]{babel}

\usepackage{graphicx}
\usepackage{dcolumn}
\usepackage{eucal}
\usepackage[dvips]{epsfig}
\usepackage{amssymb}
\usepackage{amsmath}

\begin{document}


\newcommand{\bs}{\boldsymbol}
\newcommand{\mbb}{\mathbb}
\newcommand{\mcal}{\mathcal}
\newcommand{\mfr}{\mathfrak}
\newcommand{\mrm}{\mathrm}

\newcommand{\ovl}{\overline}
\newcommand{\p}{\partial}

\renewcommand{\d}{\mrm{d}}
\newcommand{\lap}{\triangle}

\newcommand{\lan}{\bigl\langle}
\newcommand{\ran}{\bigl\rangle}

\newcommand{\bse}{\begin{subequations}}
\newcommand{\ese}{\end{subequations}}

\newcommand{\be}{\begin{eqnarray}}
\newcommand{\ee}{\end{eqnarray}}

\newcommand{\ga}{\alpha}
\newcommand{\gb}{\beta}
\newcommand{\gc}{\gamma}
\newcommand{\gd}{\delta}
\newcommand{\gr}{\rho}
\newcommand{\eps}{\epsilon}
\newcommand{\veps}{\varepsilon}
\newcommand{\gs}{\sigma}
\newcommand{\gf}{\varphi}
\newcommand{\go}{\omega}
\newcommand{\gl}{\lambda}

\renewcommand{\l}{\left}
\renewcommand{\r}{\right}

\author{A.M. Ignatov$^{1)}$, S.A. Trigger$^{2)}$}
\title{On equilibrium radiation and zero-point fluctuations in non-relativistic electron gas}

\address{
$^{1)}$ Prokhorov General Physics Institute, Russian Academy of Sciences,
Vavilova str. 38, Moscow, 119991 Russia
\\
$^{2)}$ Joint Institute for High Temperatures, Russian Academy of Sciences, Moscow 125412, Izhorskaya str. 13, build. 2, Russia\\
е-mail: satron@mail.ru}

\begin{abstract}

Examination of equilibrium radiation in plasma media shows that the spectral
the energy distribution of such radiation is different from the Planck equilibrium radiation. Using the previously obtained general relations
for the spectral energy density of equilibrium radiation in a system of charged particles, we consider radiation in an electron in the limiting case of an infinitesimal damping. It is shown that zero vacuum fluctuations which are part of the full spectral
energy distribution should be renormalized. In this case, the renormalized zero vacuum fluctuations depend on the electron density. A similar effect should exist in the general case of a quasineutral plasma.

\end{abstract}

\maketitle

Equilibrium photon energy distribution was established by Max Planck [1,2] and is a fundamental relationship,
initiated the development of quantum theory. According to Planck’s law, the spectral energy density of equilibrium radiation (SEDER), or the so-called black body radiation, defined by the expression:
\begin{eqnarray}
e_P(\omega)\equiv\frac {d E(\omega)}{d\omega} = \frac{V\hbar}{\pi^2 c^3}\frac{\omega^3}{\exp(\hbar\omega/T) -1}, \label{F1}
\end{eqnarray}
where $ V $ is the volume in which the radiation is enclosed, $ T $ is the temperature of the medium (in energy units) surrounding this volume, $ c $ is the speed of light in vacuum.
When deriving the Planck distribution, it is assumed that the photons are in thermodynamic equilibrium with the surrounding substance, although the explicit interaction of the photons with the walls of the volume is not considered. Thus, the presence of matter in the cavity where the radiation is contained should be assumed. The interaction between photons and matter in the cavity to obtain the Planck type of SEDER should be small enough to ensure the ideal photon gas and the absence of significant absorption and damping of electromagnetic radiation in the volume of $ V $ (the interaction of photons between themselves is extremely weak). At the same time, the presence of matter and a weak interaction between matter and radiation are necessary for the existence of equilibrium of a photon gas [3].
These conditions are met with a good degree of accuracy, for example, in a rarefied plasma for any frequency,
located far from the frequencies corresponding to the absorption lines of the substance.

Planck's law has been experimentally confirmed many times, for example by pumping
laser radiation into a cavity in a substance with a small hole, and subsequent observation
radiation coming out through this hole. Spectral distribution of radiation emanating from an almost empty cavity
in the measured spectral range is close to the Planck's universal curve.

Studying the explicit effect of the presence of matter on the spectral energy density of equilibrium
radiation began only recently [4-8]. Moreover, in [6] (see also [8]) it was assumed that the substance is a completely ionized plasma in which the spectrum of transverse electromagnetic oscillations (for non-relativistic and non-degenerated electrons) is determined by the expression $ \omega = \sqrt{c^2 k^2 + \omega_p^2} $. In [6], it was supposed that the damping of oscillations is negligible, and the spatial dispersion determined by the dependence of the transverse dielectric permittivity (DP) $ \varepsilon^{tr}(k, \omega) $ on the wave vector $ k $ is absent. Such an approach actually corresponds to the Brillouin approximation for the field energy in a transparent medium (see [4, 9] and references therein).
The main features of the influence of plasma on the spectral distribution of radiation revealed in [6] are the absence of radiation for the frequency range from $ \omega = 0 $ to $ \omega = \omega_p $ and a sharp drop in the total radiation energy with an increase in the characteristic dimensionless parameter $ \hbar \omega_p / T $.

A more rigorous field-theoretical approach, based on the non-relativistic approximation for particles, but taking into account both the temporal and spatial dispersion of the transverse permittivity, was developed in [7–11]. The results of [8, 9] are based on the application of the fluctuation-dissipation theorem (FDT) and the introduction of external currents, and the results [7], [10], [11] on a quantum field approach that does not contain external sources. The results are close, but lead to some differences. Since the use of third-party sources in the problem of intrinsic SEDER in plasma raises certain questions, below we consider the expression for SEDER $e(\omega)$ in the framework of the quantum field approach [7], [10],
in which the vacuum zero oscillations $e_0(\omega) = V \hbar \omega^3 /2 \pi^2 c^3 $ in the same form as in the absence of plasma are excluded from the very beginning
\begin{eqnarray}
e(\omega) =e_P(\omega) +\Delta e(\omega), \label{F2}
\end{eqnarray}
where $ \Delta e(\omega) $ is completely determined by the transverse permittivity $ \varepsilon^{tr}(k, \omega) $
\begin{eqnarray}
\Delta e(\omega)=\frac{V\hbar \omega^3}{\pi^2 c^3}\coth\left(\frac{\hbar\omega}{2T}\right)\left[\frac{c^5}{\pi\omega}\int_0^\infty d k k^4 \frac{{ \textrm {Im}}\varepsilon^{tr}(k,\omega)}{{\mid}\omega^2\varepsilon^{tr}(k,\omega)-c^2k^2\mid^2}-\frac{1}{2}\right], \label{F3}
\end{eqnarray}

However, it is not possible to establish the positivity of expression (1) at all frequencies. In this regard, two possible scenarios arise.
Either it should be accepted that expression (1) is positive for the exact (but unknown) form of the transverse DP, as well as for some correctly chosen approximations for the transverse DP, or zero-point fluctuations should be renormalized taking into account the presence of a plasma medium. In connection with the latter, one should turn to the full spectral density form $ e^{full} (\omega) $ which, as follows from the expression for $ e_0 (\omega) $ and (2), (3) equals
\begin{eqnarray}
e^{full}(\omega)=\frac{V\hbar \omega^3}{\pi^2 c^3}\coth\left(\frac{\hbar\omega}{2T}\right)\frac{c^5}{\pi \omega}\int_0^\infty d k k^4 \frac{{\textrm {Im}}\varepsilon^{tr}(k,\omega)}{(\omega^2 {\textrm {Re}}\varepsilon^{tr}(k,\omega)-c^2 k^2)^2+\omega^4({\textrm {Im}}\varepsilon^{tr}(k,\omega))^2}
 \label{F4}
\end{eqnarray}
and is always positive.

Let us consider as an example the simplest case of a collisionless electron system in the absence of spatial dispersion. At the same time, damping is also absent, since the electrons have nowhere to transfer the momentum.
For such a model, the exact DP $ \varepsilon^{el} (\omega)$ (the longitudinal and transverse DP in this case coincide), as is known, equals
 \begin{eqnarray}
\varepsilon^{el}(\omega)=1-\frac{\omega_p^2}{\omega^2}
 \label{F5}
\end{eqnarray}
Moreover, taking into account the absence of damping and taking $ \omega^2 {\textrm {Im}} \varepsilon^{tr} (k, \omega) \rightarrow +0 $ and forming the $ \delta $ -function of the argument under the integral $ \omega^2-e^{el} (\omega)-c^2 k^2 $ after integration we arrive at the expression
 \begin{eqnarray}
e^{full}(\omega)=\frac{V\hbar \omega^3}{2\pi^2 c^3}\left(1-\frac{\omega_p^2}{\omega^2}\right)^{3/2}\coth\left(\frac{\hbar\omega}{2T}\right)\theta(\omega-\omega_p)=\qquad \qquad\nonumber\\ \frac{V\hbar \omega^3}{2\pi^2 c^3}\left(1-\frac{\omega_p^2}{\omega^2}\right)^{3/2}\theta(\omega-\omega_p)+ \frac{V\hbar \omega^3}{\pi^2 c^3}\frac{1}{\exp(\hbar\omega/T)-1}\left(1-\frac{\omega_p^2}{\omega^2}\right)^{3/2}\theta(\omega-\omega_p),
 \label{F6}
\end{eqnarray}
where the first term in the second line of (6) can be considered as the renormalized vacuum fluctuations, and the second as the Planck distribution modified by particles.

Moreover, the modified vacuum fluctuations, as well as the modified Planck distribution, are already functions of the electron density. The existence of a dependence of zero-point fluctuations in a plasma medium is in a certain sense similar to the Casimir effect [12], but in the problem under consideration, the distortion of the density of vacuum photons occurs not because of the presence of surrounding walls, but because of the presence of a plasma medium.

In expression (6), the modified Planck distribution is similar to the result obtained from physical considerations in [6], [8] with the difference that in [6] (and in [8] in the corresponding limit) there is a square root of DP (5) , which characterizes the difference in the considered field approach and the approximation corresponding to FDT. However, in the latter case, the main statement of this letter on the occurrence of modified by plasma medium vacuum oscillations is valid.

As easy to show if the zero fluctuations are pick out in a vacuum form, which makes it possible to consider them as an unobservable "background",\, then the remaining "observable"\, part of  $e(\omega)\equiv\overline{e} (\omega) $\, is always negative for large values of $ \omega $, which is impossible from physical reasons. This means that the renormalization of the vacuum fluctuations in plasma is necessary for the considered model of electron gas.

A more extended consideration as well as the manifestation of forces associated with renormalized zero vacuum oscillations, will be done in a separate publication.

\end{document}